\documentclass[letterpaper,twoside]{article}
\usepackage{amssymb}
\usepackage{amsmath}
\usepackage{latexsym}
\usepackage{verbatim}
\usepackage{epsfig}
\usepackage{fancyheadings}
\usepackage{pstricks,pst-node}
\usepackage{rotating,graphicx}
\usepackage[english]{babel}

\newcommand{\papertitle}{%
Center of mass in special and general relativity and its role in an effective description of spacetime}
\newcommand{\runningtitle}{%
Relativistic center of mass: an effective description of spacetime}
\newcommand{\pauthor}{%
C.{} Chryssomalakos$^1$, H.{} Hernandez-Coronado$^1$ and E.{} Okon$^2$%
}
\newcommand{\paperauthor}{%
C.{} Chryssomalakos, H.{} Hernandez-Coronado and E.{} Okon%
}
\begin{document}
\begin{titlepage}
\vspace*{-1cm}
\begin{flushright}
\textsf{}
\\
\mbox{}
\\
\textsf{\today}
\\[3cm]
\end{flushright}
\renewcommand{\thefootnote}{\fnsymbol{footnote}}
\begin{LARGE}
\bfseries{\sffamily \papertitle}
\end{LARGE}

\noindent \rule{\textwidth}{.6mm}

\vspace*{1.6cm}

\noindent \begin{large}%
\textsf{\bfseries%
\pauthor
}
\end{large}


\phantom{XX}
\begin{minipage}{1\textwidth}
$^1$\begin{it}
\noindent  Instituto de Ciencias Nucleares \\
Universidad Nacional Aut\'onoma de M\'exico\\
Apdo. Postal 70-543, 04510 M\'exico, D.F., M\'EXICO \\[.5mm]
\end{it}

$^2$\begin{it}
\noindent  Department of Applied Mathematics\\
University of Waterloo\\
Ontario N2L 3G1, Canada\\[.5mm]
\end{it}

\texttt{chryss@nucleares.unam.mx, hcoronado@nucleares.unam.mx, eokon@uwaterloo.ca
\phantom{X}}
\end{minipage}
\\

\vspace*{3cm}
\noindent
\textsc{\large Abstract: }
In this contribution, we suggest the approach that geometric concepts ought to be defined in terms of physical operations involving quantum matter. In this way it is expected that some (presumably nocive) idealizations lying deep within the roots of the notion of spacetime might be excluded. In particular, we consider that spacetime can be probed only with physical (and therefore extended) particles, which can be effectively described by coordinates that fail to commute by a term proportional to the spin of the particles.
\end{titlepage}
\setcounter{footnote}{0}
\renewcommand{\thefootnote}{\arabic{footnote}}
\setcounter{page}{2}
\noindent \rule{\textwidth}{.5mm}

\tableofcontents

\noindent \rule{\textwidth}{.5mm}

\section{Introduction}

The only physical way that information can be extracted from spacetime is through matter, which obeys the laws of quantum mechanics. However, some of the notions that general relativity is based on are incompatible with quantum mechanics. Then, the spacetime properties related to the notions referred to above cannot be accessed by physical means.

That is, on one hand, general relativity assumes that spacetime is a differential manifold and so the notions of {\it point} and {\it path} naturally arise. Moreover, physical meaning can be assigned to those notions, for example by postulating that pointlike test particles follow geodesics. On the other hand, paths cannot be associated to quantum particles, since a particle following a path has position and momentum simultaneously defined with infinite precision, and that violates the uncertainty principle. Furthermore, even if the concept of point is not forbidden by quantum mechanics, in general, quantum states are not localized -- in spacetime -- and therefore, the concept of point becomes irrelevant.

Thus, while probing spacetime with classical pointlike test particles allows us to use paths and points, doing so with quantum test particles requires considering a generalization of points and paths. A first approach that can be taken in order to consider this, is to probe spacetime with extended classical distributions, which cannot be localized nor associated to a single path, either. We expect that the above mentioned approach can teach us ``something quantum'' about spacetime.

In the present work we explore the possibility of an effective description of spacetime. That is, we consider the possibility of obtaining a picture of spacetime in terms of properties we are used to, as points and paths, and yet, taking into account that test particles are quantum, and consequently have finite extension. In order to do so, extended test particles must be related to a single point, which should be selected according to its properties (the ``center of mass'', for instance). 

The structure of the paper is as follows: in section \ref{SR} we discuss what we think is the most suitable center of mass definition in Special Relativity and examine its properties following Pryce \cite{Pry}, in particular we find that its coordinates fail to commute in the context of Poisson brackets. In section \ref{FD} we comment on the possible implications of the noncommutativity found in section \ref{SR}, mention some further directions that can be pursued, and conclude with a comment on the center of mass definition in the General Relativity case.

\section{Center of Mass in Special Relativity}\label{SR}

Among all the possible points that can be used for effectively describing the position of an extended object, a natural choice is the center of mass, nevertheless, the problem of defining the relativistic analogue of the Newtonian center of mass for a distribution is not fully resolved. A good review of the various proposals up to 1948 is given by Pryce \cite{Pry}. In particular, for our purposes, we think that the most suitable one is such that its coordinates, $X^{\mu}$, are defined as:
\begin{equation}
X^{\mu}=\frac{J^{\mu\nu}P_\nu}{m^2}+ \alpha P^\mu,\label{CMdef}
\end{equation}
where $\mu=0,1,2,3$, $\alpha$ parameterizes the center of mass worldline, $P^\mu$ and $J^{\mu\nu}$ are the total linear and angular momenta of the mass distribution, respectively, and $m^2=P^\mu P_\mu$. If we think of the mass distribution as a set of pointlike particles, definition (\ref{CMdef}) is the {\it average of particle positions, weighted by their total energies in the center of momentum frame, i.e., the frame in which the total three momentum vanishes}. According to the above definition, the center of mass satisfies the following properties:
\begin{enumerate}
 \item [a.] Its spatial components are part of a four vector, whose zeroth component is the time at which they are measured.

 \item [b.] It is at rest in the ``center of momentum'' frame.

 \item [c.] When no external forces act on the distribution, it moves with constant velocity $P^\mu/P^0$.

\end{enumerate}

By looking at the above properties, it could be concluded that nothing really gets modified by considering the extension of test particles while probing spacetime, however, it is interesting to note that the spatial components of the center of mass defined as (\ref{CMdef}) satisfy the nontrivial commutation relations\footnote{It is precisely because of this fact that Pryce concluded that there was not a single fully satisfactory relativistic center of mass definition.} (in the sense of Poisson brackets),
\begin{equation}
 [X^i,X^j]=\epsilon^{ij}_{\;\;\;k}\left(\frac{S^k}{E^2}+\frac{\mathbf{S}\cdot\mathbf{P}}{m^2 E^2}\;P^k\right),\label{Comm}
\end{equation}
with $\mathbf{S}=(S^{23},S^{31},S^{12})$ where $S^{\mu\nu}=J^{\mu\nu}-X^{\mu} P^{\nu}+X^{\nu} P^{\mu}$. The (classical) noncommutativity of the $X$'s found in (\ref{Comm}) seems to follow from the fact that the system is extended in space. In order to clarify the last statement, suppose the system is observed in the center of momentum frame, where it is found to have non zero angular momentum along the $z$-axis, and therefore non-zero $Sz$ (since angular momentum in the center of momentum frame is spin by definition). Then upon quantization and according to the Heisenberg uncertainty principle, the above expression implies that the system's center of mass must be situated within a fuzzy region of the x-y plane for any point on the z-axis. This should be compared to the Newtonian case, where the center of mass can be located exactly since their coordinates commute. This can be translated into the fact that the behavior of extended systems reduces exactly to that of a pointlike particle located at its center of mass (modulo its internal motion), only in the Newtonian limit. The uncertainty exhibited by the relativistic center of mass position shows that its ``mean position'' cannot be defined sharply, and this is exclusively a relativistic effect. 

The relations (\ref{Comm}) become relevant in the quantum version, where the center of mass coordinates are elevated to operators and the Poisson brackets to commutators. Their relevance in the quantum gravitational context, resides in the aspiration of finding a quantum analogue to the classical geodesic motion of point particles. As discussed earlier, the hope is to assign some sort of effective path to a quantum particle, for which a mean position is needed because its state is, in general, spread out. Under the most favorable circumstances, given (\ref{Comm}), the quantum particle, described by an effective point particle located at the ``mean position'', will feel some sort of average of the metric over a region whose extent is of the order of the r.h.s. of (\ref{Comm}).

Notice also that the present noncommutativity differs from others in the literature, in the sense that it naturally arises from physical considerations, on one side, and in that it depends on the particle properties so that the physical interpretation is fundamentally different, on the other (for a more extensive discussion the reader is referred to \cite{Chr1}). It is also interesting to note that this same kind of noncommutativity appears in several different physical contexts ({\it e.g.}, \cite{Cas}, \cite{Bet}).	

Another feature of (\ref{CMdef}) worth-mentioning is that it forces us to consider spin, which is included in the total angular momentum $J^{\mu\nu}$. In the context of classical systems, as defined above, spin means {\it intrinsic angular momentum}, {\it i.e.}, the part of the angular momentum which does not depend on the origin of coordinates. If spin is not taken into account, the definition turns out to be nonassociative, that is, when computing the center of mass for, say, three mass distributions, the result depends on the order in which they are composed. This occurs even if the initial distributions have spin zero because the mass distribution resulting from such composition has in general non-zero spin. 

Finally, note that the position coordinates of a single spinless particle commute, and expression (\ref{Comm}) then establishes that (a quantum version of) it can be exactly localized, as in standard quantum mechanics. Notice likewise that it implies for an electron's center of mass to be found within a region whose area is of the order of (Compton wavelength)$^2\sim10^{-25}m^2$. This is, clearly, a consequence of considering only one-particle states in our description, since, as it is well known, in Dirac's theory for spin one-half particles -- where antiparticle states are admitted -- the position operators are commutative. Invocation of antiparticles, miraculously, restores commutativity.

\section{Further directions}\label{FD}

Let us try to understand what we have learned from the previous discussion. The starting point was the assumption that the only way in which we can probe spacetime is through matter, which is quantum, and is described by a nonlocalized, in general, wavefunction. Intending to take account of the nonlocality of quantum states into the description of spacetime, classical extended matter was considered. In the Newtonian limit, classical extended matter can be described exactly as an effective point particle, located at the center of mass, with all of its classical properties. In the relativistic situation, its effective description naturally leads to noncommutative center of mass coordinates (in the sense of Poisson brackets). And finally, quantum mechanics has to be called for assigning a nonlocality interpretation to the noncommutative coordinates. This suggests that noncommutativity could be an important element for considering ``quantum aspects'' of spacetime.

An immediate next step in the program, should be to look for some sort of effective metric such that ``effective quantum matter'' follow effective geodesics (conveniently defined). Of course, it is not clear that such effective entities should exist.

Another direct way to proceed is to consider general spacetimes, that is, to consider General Relativity, where the situation is much more complicated. The authors know of at least three inequivalent center of mass definitions (\cite{Bei}-\cite{Sud}) that reduce to (\ref{CMdef}) when spacetime is flat. They have different properties and the choice of which should be used, depends on the specific physical situation that is dealt with. The differences between them stem from the fact that in curved spacetimes, parallel transport is not trivial and then the comparison between vectors associated to different points of the distribution is nontrivial. As a consequence, there is not only one way to define the total mass or total angular and linear momenta. Explicit expressions for the previous definitions are complicated and, as expected, expressions for the commutators have not been computed. The only effort towards that direction, to the knowledge of the authors, is the work of Bonder and Sudarsky, who are working nowadays on the computation of the Poisson brackets for the coordinates of their center of mass definition.

\section*{Acknowledgments}
CC and HH would like to thank the organizers of DICE 2008, in particular, Thomas Elze, for efficiently handling their requests and the high scientific level they helped to maintain. CC and HH also wish to acknowledge partial financial support from DGAPA-UNAM projects IN 121306.


\vspace{1cm}
\begin{thebibliography}{10}

\bibitem{Pry} Pryce M H L 1948, {\it Proc. Royal Soc. Lon.} \textbf{195} 62-81

\bibitem{Chr1} Chryssomalakos C, Hern\'andez H, Okon E and V\'azquez-Montejo P 2007, {\it J. Phys.: Conf. Series} \textbf{68} 012003

\bibitem{Cas} Casalbouni R 1976 {\it Nuovo Cimento} A \textbf{33} 389

\bibitem{Bet} Bette A 1984 {\it J. Math. Phys.} \textbf{25} 2456

\bibitem{Bei} Beiglb\"ock W 1967 {\it Commun. Math. Phys.} \textbf{5} 106.

\bibitem{Dix} Dixon W G 1964 {\it Nuovo Cimento} \textbf{34} 317. W.G.
Dixon 1970 \textit{Proc. Roy. Soc. London A} \textbf{314} 499.

\bibitem{Sud} Bonder Y and Sudarsky D {\it Private communication.}

\end{thebibliography}
\end{document}